\documentclass[10.5pt,onecolumn,pdflatex,sn-mathphys-num]{jicn}

\usepackage{graphicx}%
\usepackage{multirow}%
\usepackage{amsmath,amsthm,amssymb,amsfonts,dsfont,pifont,bm,bbm,mathrsfs,mathtools,nicefrac,extarrows,relsize}%
\usepackage{mathrsfs}%
\usepackage[title]{appendix}%
\usepackage{textcomp}%
\usepackage{manyfoot}%
\usepackage{algorithm}%
\usepackage{algorithmicx}%
\usepackage{algpseudocode}%
\usepackage{listings}%
\usepackage{overpic}
\usepackage{wrapfig}
\usepackage{subcaption}
\usepackage[dvipsnames,table,xcdraw]{xcolor}
\usepackage{colortbl}
\usepackage{adjustbox,diagbox, tabularray}
\usepackage{multicol} 
\usepackage{lipsum}
\usepackage{threeparttable}
\usepackage[utf8]{inputenc}
\usepackage{booktabs}
\usepackage{caption}
\usepackage{titlesec}
\usepackage{tablefootnote}
\usepackage{times}
\usepackage{mathptmx}
\usepackage{float}
\usepackage{arydshln}
\usepackage[english]{babel}
\usepackage{tabularx}
\addto\captionsenglish{\renewcommand{\figurename}{Figure}}
\usepackage{setspace}
\def\JournalName{Journal of Intelligent Computing and Networking} 
\def\JournalURL{https://www.ffspub.com/index.php/jicn/index} 
\def\JournalEmail{jicn.office@ffspub.com} 
\def\JournalISSNprint{ 3079-9228 (print)} 
\def\JournalLogoFile{jicnlogo.png} 

\def\authorshort{YU et al.} 

\def\PaperDOI{10.64509/jicn.11.xx} 

\def\PaperCopyrightOwner{2025 The authors. This article is an open access article distributed under the terms and conditions of the Creative Commons Attribution (CC BY) license (https://creativecommons.org/licenses/by/4.0/).}

\usepackage{geometry}
\usepackage{fancyhdr}
\usepackage[hang,flushmargin]{footmisc}

\titleformat{\section}
  {\normalfont\fontsize{16}{16}\bfseries}
  {\thesection}
  {1em}
  {} 
  \titlespacing{\section}{0pt}{1em}{1em} 
\titleformat{\subsection}
  {\normalfont\fontsize{12}{12}\bfseries}
  {\thesubsection}
  {1em}
  {}
  \titlespacing{\subsection}{0pt}{0.8em}{0.8em} 
\titleformat{\subsubsection}
  {\normalfont\fontsize{10}{10}\bfseries}
  {\thesubsubsection}
  {1em}
  {}
  \titlespacing{\subsubsection}{0pt}{0.5em}{0.5em}

\def\journalinfohead{\authorshort{} 
}
\makeatletter
\def\blfootnote#1{\gdef\@thefnmark{}\@footnotetext{#1}}
\makeatother

\renewcommand{\footnoterule}{\kern -3pt \hrule width 0.3\columnwidth \kern 2.6pt}

\hypersetup{
	colorlinks=true,
	urlcolor=blue,
	citecolor=blue,
	linkcolor=blue
}

\theoremstyle{thmstyleone}%
\theoremstyle{thmstyletwo}%

\theoremstyle{thmstylethree}%

\definecolor{NearOOD}{HTML}{E6ECE3}
\definecolor{FarOOD}{HTML}{ffefe0}

\begin{document}

\noindent
\begin{minipage}[c][2.5cm][c]{0.75\textwidth}
  \fontsize{9}{11}\selectfont
  \raggedright
  \textbf{\JournalName} \\
  \JournalURL \\
  ISSN \JournalISSNprint \\
  E-mail: \JournalEmail
\end{minipage}%
\hfill
\begin{minipage}[c][2.5cm][c]{0.2\textwidth}
  \flushright
  \includegraphics[width=3.5cm]{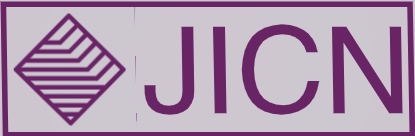}
\end{minipage}
\fontsize{10}{12}\selectfont
\noindent\textit{Article}
\begin{center}
  \fontsize{20}{24}\bfseries
  GraphPilot: GUI Task Automation with One-Step LLM Reasoning Powered by Knowledge Graph
\end{center}

\begin{center}
  \fontsize{11}{14}\selectfont
  Mingxian YU\textsuperscript{1},
  Siqi LUO\textsuperscript{1},
  Xu CHEN\textsuperscript{1,\textdagger}
  \\
  \fontsize{9}{11}\selectfont\itshape
  \textsuperscript{1}School of Computer Science and Engineering, Sun Yat-sen University, Guangzhou 510006, China\\

  \textsuperscript{\textdagger}E-mail: chenxu35@mail.sysu.edu.cn
  \\
  \vspace{1em}
\end{center}
\vspace{1em}

\noindent\textbf{Abstract:} Mobile graphical user interface (GUI) agents are designed to automate everyday tasks on smartphones. Recent advances in large language models (LLMs) have significantly enhanced the capabilities of mobile GUI agents. However, most LLM-powered mobile GUI agents operate in stepwise query-act loops, which incur high latency due to repeated LLM queries. We present GraphPilot, a mobile GUI agent that leverages \textit{knowledge graphs} of the target apps to complete user tasks in almost \textit{one LLM query}.
GraphPilot operates in two complementary phases to enable efficient and reliable LLM-powered GUI task automation. In the offline phase, it explores target apps, records and analyzes interaction history, and constructs an app-specific knowledge graph that encodes functions of pages and elements as well as transition rules for each app. In the online phase, given an app and a user task, it leverages the knowledge graph of the given app to guide the reasoning process of LLM. When the reasoning process encounters uncertainty, GraphPilot dynamically requests the HTML representation of the current interface to refine subsequent reasoning. Finally, a validator checks the generated sequence of actions against the transition rules in the knowledge graph, performing iterative corrections to ensure it is valid. The structured, informative information in the knowledge graph allows the LLM to plan the complete sequence of actions required to complete the user task.
On the DroidTask benchmark, GraphPilot improves task completion rate over Mind2Web and AutoDroid, while substantially reducing latency and the number of LLM queries.
\vspace{1em}

\vspace{8mm}

\blfootnote{\textsuperscript{\textdagger} Corresponding author: Xu Chen}
\blfootnote{\textcopyright\ \PaperCopyrightOwner}

\begin{multicols}{2}

  \section{Introduction}

Mobile graphical user interface (GUI) agents have emerged as a promising paradigm for automating smartphone tasks through natural language instructions. Powered by recent advances in large language models (LLMs), these agents can understand user intents and interact with apps in a generalizable way. In contrast to traditional solutions that require predefined scripts for each app and scenario, LLM-powered mobile GUI agents leverage the understanding and reasoning abilities of LLMs to interact dynamically across different interfaces~\cite {liu2025llm}.

Despite these advances, most contemporary LLM-powered mobile GUI agents follow a stepwise interaction paradigm. In this approach, the agent iterates through query-act loops: at each interface, it queries the LLM with the current GUI (often a screenshot~\cite{zhang2025appagent} or view hierarchy~\cite{wen2024autodroid}) and the task description, receives a single action in return (such as ``click on button X''), performs that action, and then repeats for the next interface until task completion. A simple example of this process is shown in Figure~\ref{fig:stepwise-example}.
Although straightforward, this iterative process is highly inefficient, as repeated LLM queries at each interaction accumulate substantial latency. Such a stepwise design inherently trades computational efficiency for incremental reasoning, leading to considerable overhead in multi-step tasks.

Currently, GUI agents such as Mind2web~\cite{deng2023mind2web} and AutoDroid~\cite{wen2024autodroid} must issue multiple LLM queries per user task, leading to significant network latency. So we introduce GraphPilot, a mobile GUI agent that improves task completion rate while reducing latency by almost one LLM query with the help of knowledge graphs. First, GraphPilot constructs app-specific \textit{knowledge graphs} that encode the functions of pages and elements, as well as the transition rules about element-to-page transitions for each app. Such a representation of a knowledge graph naturally captures the complex relationships between pages and elements, providing structured contextual knowledge for reasoning. Second, GraphPilot enables almost \textit{one LLM query} to generate a complete sequence of actions for a user task by leveraging the previously constructed knowledge graph, thereby significantly reducing the need for repetitive query–act loops.

\begin{center}
    \includegraphics[width=\linewidth]{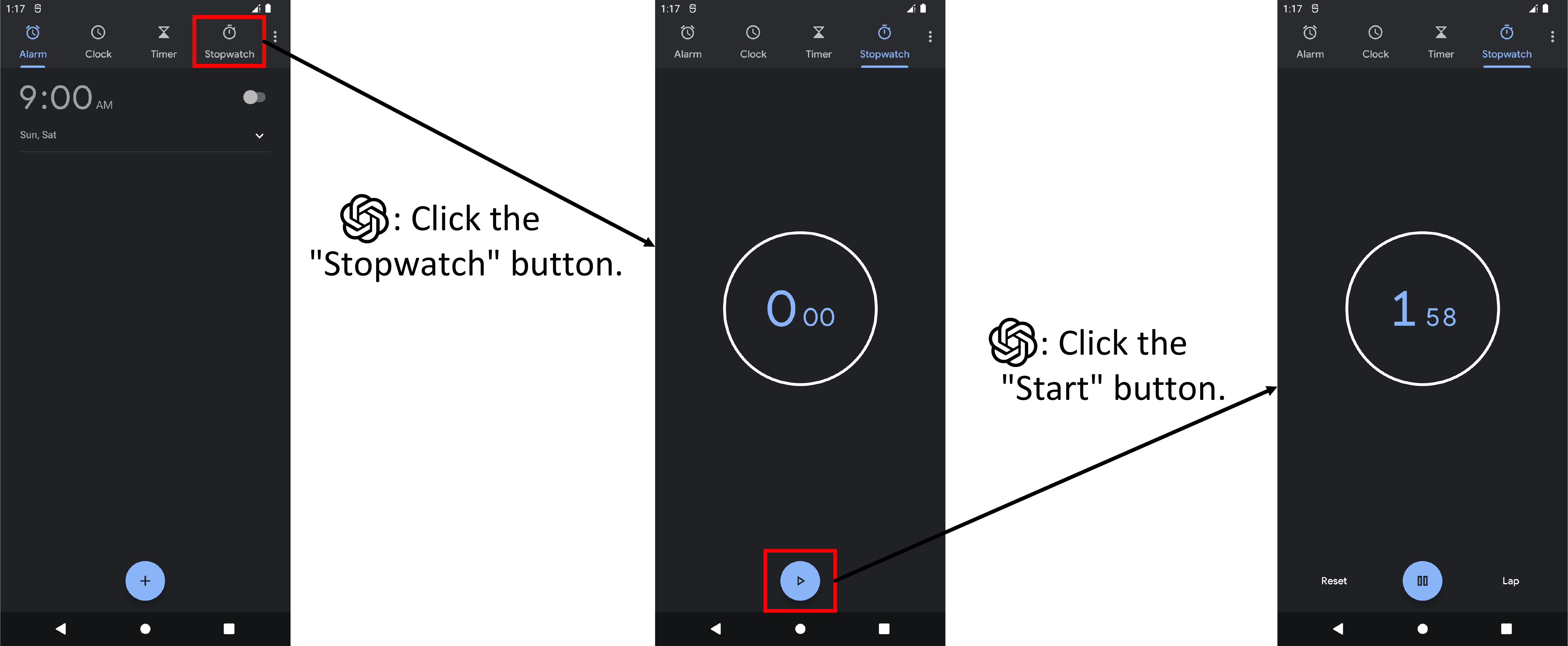}
    \captionof{figure}{A simple example of the process of the stepwise GUI agents handling the task ``start the stopwatch'' in the Clock app.}
    \label{fig:stepwise-example}
\end{center}

GraphPilot operates in two phases. In the offline phase, it explores target apps, records exploration history, and extracts the functions of pages and elements, along with their transition rules, by querying the LLM. The extracted knowledge is then organized into app-specific knowledge graphs, which serve as a structured representation of the app interaction logic. In the online phase, GraphPilot leverages this knowledge graph to efficiently generate a complete sequence of actions for a given app and a user task with almost one LLM query. When uncertainty arises during reasoning, GraphPilot dynamically extracts the current HTML representation of the interface to guide subsequent reasoning. Finally, a validator checks the generated sequence of actions against the transition rules encoded in the knowledge graph, ensuring it is valid. By integrating a structured app-specific knowledge graph into LLM reasoning, GraphPilot bridges static understanding and dynamic task execution, achieving both higher efficiency and reliability in mobile GUI automation.

\textbf{Essence:} In the stepwise paradigm, each LLM query contains only limited contextual information. This requires multiple incremental queries to process user tasks step by step, which increases network latency. In contrast, GraphPilot incorporates rich contextual information into LLM queries, enabling the LLM to generate a more complete sequence of actions. This approach significantly reduces the number of LLM queries, effectively lowering network latency.

Experiments on DroidTask demonstrate that GraphPilot outperforms both Mind2Web and AutoDroid. It achieves the highest task completion rate of 74.1\%, compared to 65.2\% for Mind2Web and 62.0\% for AutoDroid, while reducing the latency of LLM queries by 70.4\% and 66.7\%, respectively.

Our main contributions are summarized as follows:
\begin{enumerate}
    \item \textbf{The construction of app-specific knowledge graphs.} GraphPilot constructs a knowledge graph for each app, encoding page and element functions, as well as transition rules. This structured representation enables a comprehensive understanding of the app's interaction logic, providing accurate, informative knowledge for task execution.
    \item \textbf{Almost one LLM query to generate a complete sequence of actions.} GraphPilot departs from the prevalent stepwise query–act paradigm, which suffers from high latency due to repeated LLM queries. By leveraging the informative knowledge graph for each app, GraphPilot generates a complete sequence of actions in almost one LLM query, substantially reducing the latency of LLM queries while maintaining reasoning quality.
\end{enumerate}
\section{Related Work}

Traditional mobile GUI task automation has predominantly relied on scripted approaches and Robotic Process Automation (RPA) tools such as Monkey~\cite{monkey} and Sikuli~\cite{sikulix}, which execute predefined sequences of GUI actions. However, these conventional methods suffer from brittleness, necessitating manual maintenance whenever interface layouts evolve~\cite{liu2025llm}. In contrast, recent advances in large language model (LLM)-powered agents demonstrate the capability to interpret high-level natural language instructions and dynamically adapt to evolving application interfaces. This section surveys the key developments in LLM-powered mobile GUI automation.

\subsection{Interface Representation}

Effective GUI automation requires a comprehensive representation of interface elements and visual content. Contemporary approaches can be broadly categorized into two primary representation paradigms: text-based and multimodal (text and image) categories. Text-based methods leverage structured GUI data, including accessibility trees and Document Object Model (DOM)-like structures, as demonstrated in DroidBot-GPT~\cite{wen2023droidbot} and AutoDroid~\cite{wen2024autodroid}. Conversely, multimodal approaches integrate visual analysis through Vision-Language Models (VLMs): ScreenAI~\cite{baechler2024screenai} represents a dedicated vision-language architecture for screenshot interpretation, while VisionDroid~\cite{liu2024vision} and GUI Narrator~\cite{wu2024gui} augment textual input with OCR-extracted text and icon annotations via Set-of-Mark (SoM) prompting techniques. Advanced multimodal agents such as Mobile-Agent-v2~\cite{wang2024mobile} and OmniParser~\cite{wan2024omniparser} incorporate sophisticated icon recognition and layout analysis capabilities.
Our proposed GraphPilot adopts a text-based paradigm via an HTML-based interface representation, achieving high computational efficiency while preserving semantic richness and avoiding visual processing overhead.

\subsection{Knowledge and Memory Representations}

To handle diverse app functionality, recent agents augment LLMs with memory or knowledge structures. AutoDroid~\cite{wen2024autodroid} builds a GUI Transition Graph (UTG) by exploring the app and injects this app-specific memory into prompts. MobileGPT~\cite{lee2024mobilegpt} stores a hierarchical memory of tasks, subtasks, and actions for recall. AppAgent~\cite{zhang2025appagent} uses autonomous exploration and human demonstrations to build a multimodal \emph{knowledge base} for each app. RAG-based updating appears in AppAgentX~\cite{jiang2025appagentx} and AdaptAgent~\cite{verma2024adaptagent}. These approaches show the benefit of app-level context.
While GraphPilot is novel in organizing app-specific knowledge into knowledge graphs that capture both the behaviors and structures of apps, enabling the reasoning process in nearly a single LLM query.



\subsection{Task Planning and Action Execution}

Most LLM-driven GUI agents operate in an iterative query-act cycle, where a single agent repeatedly observes the interface state and then determines the next action. For example, DroidBot-GPT~\cite{wen2023droidbot} and AutoDroid~\cite{wen2024autodroid} both use an LLM to choose the next UI element to interact with at each step, continuing this loop until the task is completed. This stepwise strategy offers adaptability to dynamic interfaces but incurs the overhead of multiple LLM queries. To improve efficiency, MobileGPT~\cite{lee2024mobilegpt} introduces a hierarchical ``explore-select-derive-recall'' workflow and a multi-level memory (tracking tasks, subtasks, and actions) that significantly boosts automation accuracy and speed through better context management. Likewise, CogAgent~\cite{hong2024cogagent} follows a similar iterative paradigm, leveraging a visual-language model to reason about the GUI for each incremental action.

To handle more complex tasks, some agents focus on making plans described in natural language executable on real interfaces or even employ multiple specialized agents. SeeAct~\cite{zheng2024gpt} first drafts a tentative sequence of actions, then performs action grounding and element alignment to locate the correct target element and translate each step into a concrete click or input action. The emphasis is on this grounding and alignment procedure, which reliably turns high-level plans into runnable operations. Alternatively, multi-agent systems have been proposed to divide responsibilities among cooperative agents. For instance, MMAC-Copilot~\cite{song2024mmac} coordinates multiple modalities through role-specific agents, and MobileExperts~\cite{zhang2024mobileexperts} uses a dynamic team of tool-driven agents to accomplish tasks jointly. Other works integrate specialized modules into the agent framework. For example, ClickAgent~\cite{hoscilowicz2024clickagent} combines an MLLM-based reasoner with a dedicated GUI locator model to better pinpoint interface elements, while Ponder \& Press~\cite{wang2024ponder} adopts a divide-and-conquer design that splits the problem into high-level instruction interpretation followed by low-level visual element localization. By partitioning the task in these ways, such approaches aim to reduce error propagation and improve success rate.

Despite these advances, the majority of LLM-driven agents still rely on incremental action generation, which accumulates latency over multiple iteration cycles. In contrast, our proposed GraphPilot agent generates a complete action sequence in almost a single interaction with the LLM by leveraging knowledge graphs, rather than looping through each step individually. This one-step planning capability eliminates the need for repeated LLM queries, significantly reducing network latency.
\section{Design of GraphPilot}

This section provides a detailed design of the two-phase architecture of GraphPilot and the functions of its components.

\subsection{Overview}

\begin{figure*}[!htb]
    \centering
    \includegraphics[width=\linewidth]{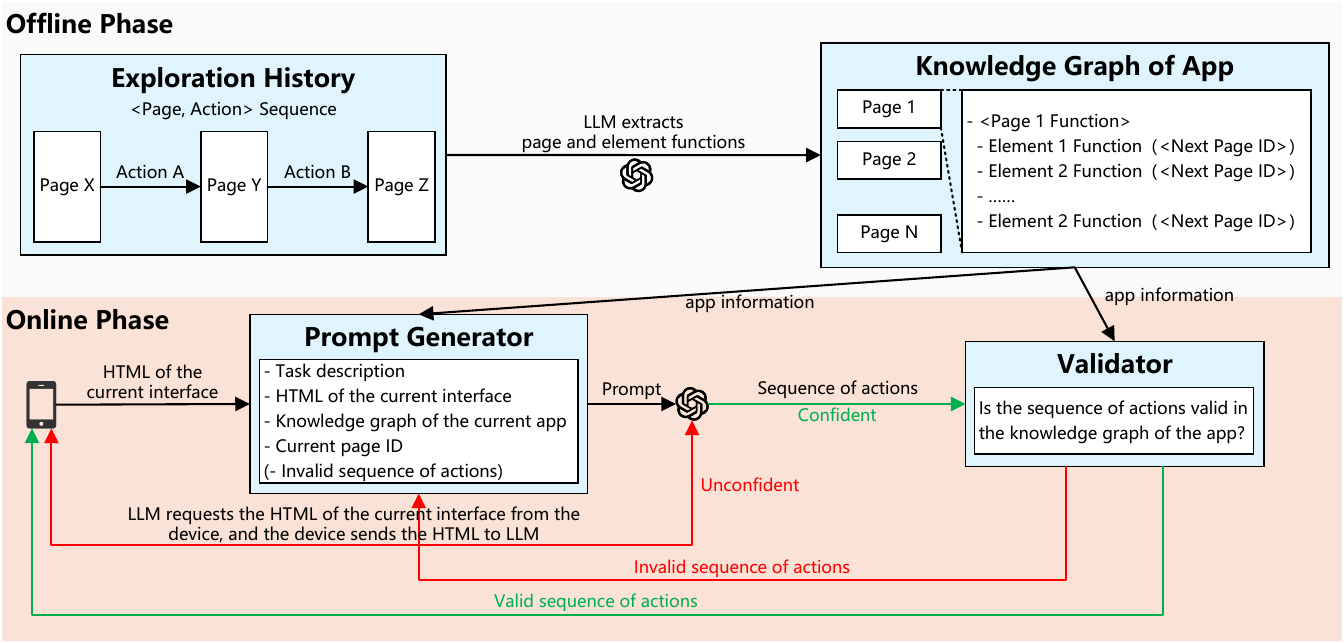}
    \caption{Overview of GraphPilot.}
    \label{fig:overview}
\end{figure*}

The overview of GraphPilot is shown in Figure~\ref{fig:overview}. GraphPilot operates in two phases: an offline phase and an online phase.

The offline phase focuses on constructing knowledge graphs for each app. These knowledge graphs provide enough and accurate information for the online phase of GraphPilot. During this phase, GraphPilot first explores mobile apps to build a comprehensive understanding of their functions and structures. It maintains detailed \textit{exploration history} \ding{182} about the sequence of actions and corresponding interface changes. Then GraphPilot analyzes the functionality of each page and its constituent elements using an LLM. Additionally, GraphPilot records transition information in the history, such as ``interacting with element X on page A will lead to page B.'' This \textit{element-to-page} transition information provides directional guidance for the online phase of GraphPilot. Finally, all such information is consolidated into a \textit{knowledge graph} \normalsize\ding{183} that captures both the behaviors (functions of pages and elements) and structures (element-to-page transition rules) of the apps. An example of the knowledge graph for an app is shown in Figure~\ref{fig:kb_example}. The nodes in the knowledge graph represent the functions of pages and elements, and the edges represent the element-to-page transition rules. Finally, in the online phase, GraphPilot will use the knowledge graph as an informative context to support its task planning.

\begin{center}
    \includegraphics[width=\linewidth]{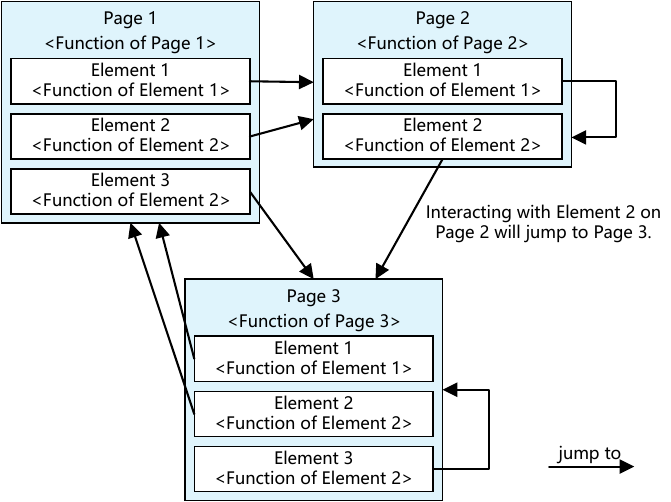}
    \captionof{figure}{An example of the knowledge graph for an app.}
    \label{fig:kb_example}
\end{center}

\begin{figure*}[!htb]
    \centering
    \includegraphics[width=\linewidth]{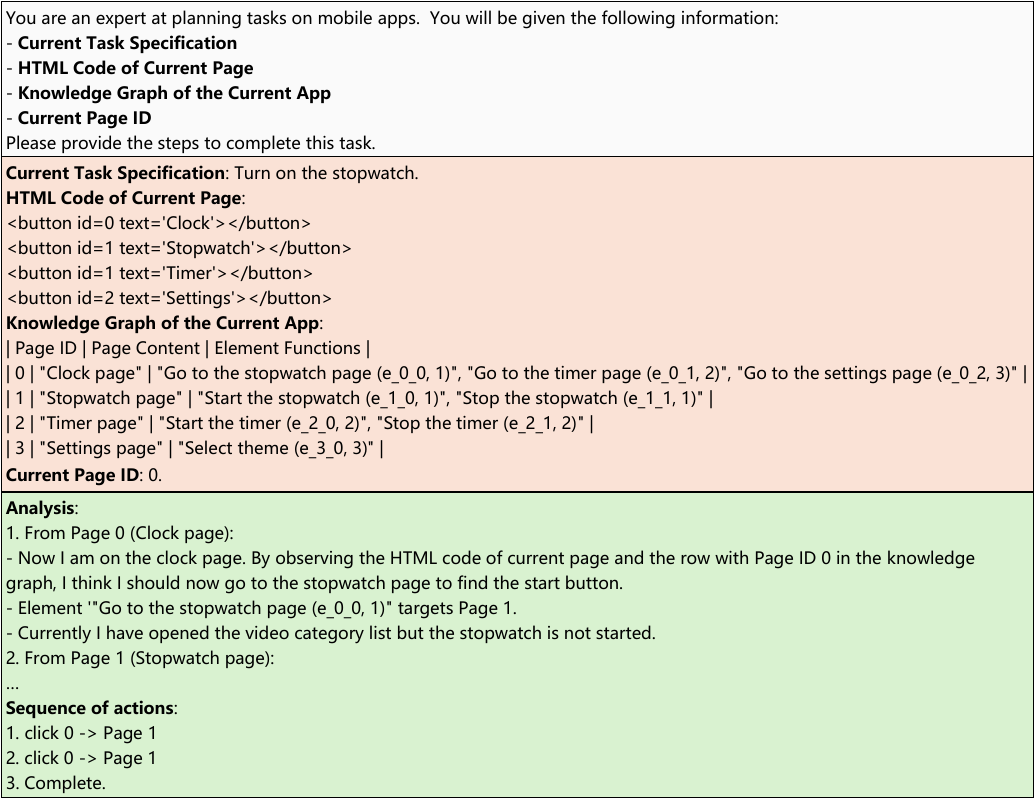}
    \caption{A simple example of the prompt generated by the prompt generator. The gray part represents a fixed background description. The red part represents dynamic content. The green part shows an example output generated by the LLM. (X, A) in the knowledge graph represents that interacting with element X will jump to page A.}
    \label{fig:example-prompt}
\end{figure*}

The online phase leverages the knowledge graph constructed in the offline phase to perform task planning. Given a user-specified task, a \textit{prompt generator} \ding{184} generates a prompt that includes the task description and the knowledge graph (erroneous actions from the validator may also be included, which will be discussed later). The generated prompt is sent to the LLM to generate a candidate sequence of actions. This approach differs from traditional stepwise GUI agents that query the LLM at each interaction step.
GraphPilot may encounter uncertainty during its reasoning. When this happens, it can dynamically request the HTML of the current interface. This provides additional context to proceed when the knowledge graph is insufficient. After GraphPilot generates a candidate sequence of actions, a \textit{validator} \ding{185} examines its correctness by checking the transition rules in the knowledge graph.
If any rule violations occur, the validator detects the specific erroneous actions and sends them back to the prompt generator for resubmission, guaranteeing that only valid sequences are executed.

This two-phase architecture enables GraphPilot to achieve both high accuracy and low latency. The offline phase ensures comprehensive coverage of behaviors and structures of apps, while the online phase enables rapid task completion through almost one LLM query with prompt refinement when necessary.

\subsection{Formulation of Offline Phase}

The offline phase constructs a knowledge graph for each app by exploring the target app, analyzing the functions of its pages and elements, and extracting element-to-page transition rules from the recorded interaction history. We formally define the knowledge graph construction process as shown in Algorithm~\ref{alg:kb}.

\textbf{App and Exploration History.} Consider a target mobile app denoted as \(A_i\), where \(i\) is the app index. During exploration, we record the exploration history \(S_i=\{(H^{(t)},s^{(t)},H^{(t+1)})\}_{t=1}^{T}\), which consists of \(T\) sequential interaction steps. Each step is represented by a tuple containing: (1) \(H^{(t)}\), the HTML representation of the current interface state at step \(t\); (2) \(s^{(t)}\), the action performed at step \(t\); and (3) \(H^{(t+1)}\), the resulting HTML representation after the action.

\textbf{Actions and Elements.} Each action \(s^{(t)}\) is decomposed as \(s^{(t)}=(a^{(t)},e^{(t)})\), where \(a^{(t)}\) represents the specific action type and \(e^{(t)}\) represents the target element at step \(t\). The action type \(a^{(t)}\) is selected from a predefined action space \(\{\textsf{click},\ \textsf{text},\ \textsf{stop}\}\).
The element \(e^{(t)}\) identifies element being interacted with on page \(p^{(t)}\), while \(e^{(t)}=0\) indicates no element interaction (used for the \textsf{stop} action).
For the terminal step, we enforce \(a^{(T)}=\textsf{stop}\) and \(e^{(T)}=0\), indicating that the exploration sequence finishes without targeting any specific element. Function \(\textsc{PageID}(\cdot)\) extracts the page ID based on the given HTML representation. When $t=1$, $p_\mathrm{curr}$ denotes the initial page ID (line 4), and no preceding page $p_\mathrm{prev}$ exists. Accordingly, we set $p_\mathrm{prev}$ to null in this case (line 3). 

\begin{center}
    \begin{algorithm}[H]
        \caption{Offline Knowledge Graph Construction for app \(A_i\)}
        \label{alg:kb}
        \begin{algorithmic}[1]
            \Require Exploration history \(S_i=\{(H^{(t)},s^{(t)},H^{(t+1)})\}_{t=1}^{T}\)
            \Ensure Knowledge graph \(K_i\)
            \State \(K_i\leftarrow \varnothing\)
            \For{\(t=1\) to \(T\)}
            \State \(p_{\mathrm{prev}}\leftarrow \textsc{PageID}(A_i,H^{(t-1)})\) (set to \(\mathrm{null}\) if \(t=1\))
            \State \(p_{\mathrm{curr}}\leftarrow \textsc{PageID}(A_i,H^{(t)})\)
            \State \(p_{\mathrm{next}}\leftarrow \textsc{PageID}(A_i,H^{(t+1)})\) (set to \(p_{\mathrm{curr}}\) if \(t=T\))
            \If{\(F^{\mathrm{page}}_{p_{\mathrm{curr}}}\) is undefined}
            \State \(F^{\mathrm{page}}_{p_{\mathrm{curr}}} \leftarrow \textsc{QueryPageFunction}(\)
            \Statex \hspace{2em} \(H^{(t-1)},s^{(t-1)},H^{(t)},s^{(t)},H^{(t+1)})\)
            \State \(K_i \leftarrow K_i \cup \{(``\mathrm{Page}",p_{\mathrm{curr}},F^{\mathrm{page}}_{p_{\mathrm{curr}}})\}\)
            \EndIf
            \State \((a^{(t)}, e^{(t)}) \leftarrow s^{(t)}\) (set \(e^{(t)}=0\) if \(t=T\))
            \If{\(e^{(t)} \ne 0\) and \(F^{\mathrm{elem}}_{p_{\mathrm{curr}},e^{(t)}}\) is undefined}
            \State \(F^{\mathrm{elem}}_{p_{\mathrm{curr}},e^{(t)}} \leftarrow \textsc{QueryElementFunction}(\)
            \Statex \hspace{2em} \(H^{(t)},s^{(t)},H^{(t+1)})\)
            \State \(K_i \leftarrow K_i \cup \{(``\mathrm{Element}",p_{\mathrm{curr}},e^{(t)},F^{\mathrm{elem}}_{p_{\mathrm{curr}},e^{(t)}})\}\)
            \State \(K_i \leftarrow K_i \cup \{(``\mathrm{Transition}",p_{\mathrm{curr}},e^{(t)},p_{\mathrm{next}})\}\)
            \EndIf
            \EndFor
            \State \Return \(K_i\)
        \end{algorithmic}
    \end{algorithm}
\end{center}

\textbf{Knowledge Graph Structure.} The constructed knowledge graph for app \(A_i\) is denoted as \(K_i\). For each page \(p_j\), the knowledge graph stores three types of information: (1) \(F^{\mathrm{page}}_{p_j}\), a natural language description of the function of page \(p_j\), which is obtained by function \(\textsc{QueryPageFunction}(\cdot)\) (line 7); (2) \(F^{\mathrm{elem}}_{p_j,e_k}\), a natural language description of the functionality of element \(e_k\) on page \(p_j\), which is obtained by function \(\textsc{QueryElementFunction}(\cdot)\) (line 12); and (3)
the tuple starting with ``Transition'' in \(K_i\) represents the transition rules, and the following three elements represent the starting page, the element interacted with, and the arrival page, respectively (line 14).


\subsection{Formulation of Online Phase}

During the online phase, GraphPilot generates a sequence of actions utilizing the knowledge graph constructed in the offline phase. GraphPilot iteratively proposes a sequence of actions and refines it according to the transition rules through the validator.
As illustrated in Figure~\ref{fig:example-prompt}, while generating the Analysis part, the LLM may reach an intermediate step where the knowledge graph no longer provides enough information to proceed. When this happens, the LLM is unconfident to continue the inference and therefore requests the HTML representation of the current interface to obtain the missing context. Conversely, if the LLM can complete the reasoning process in the Analysis part without encountering informational insufficiency, it is considered to be confident. 
This process is shown in Algorithm~\ref{alg:online}.

\begin{algorithm}[H]
    \captionof{algorithm}{Online Action Sequence Generation for app \(A_i\)}
    \label{alg:online}
    \begin{algorithmic}[1]
        \Require User task description \(d\), knowledge graph \(K_i\), current page \(p_{\mathrm{curr}}\)
        \Ensure A validated action sequence \(S\)
        \State \(S\leftarrow \varnothing\)
        \State \(I\leftarrow \varnothing\)
        \While{\(S=\varnothing\) and maximum number of iterations is not reached}
        \State \(prompt \leftarrow \textsc{PromptGenerator}(d,K_i,I,p_{\mathrm{curr}})\)
        \State \(S\leftarrow \textsc{QueryActionSequence}(prompt)\)
        \State \(I\leftarrow \textsc{Validator}(S,K_i)\)
        \If{\(I \neq \varnothing\)}
        \State \(S\leftarrow \varnothing\)
        \EndIf
        \EndWhile
        \State \Return \(S\)
    \end{algorithmic}
\end{algorithm}

Given app \(A_i\) and user task description \(d\), GraphPilot generates a sequence of actions \(S=\{(p^{(t)},a^{(t)},e^{(t)},p^{(t+1)})\}_{t=1}^T\), where each step \(t\) contains: (1) \(p^{(t)}\), the page at step \(t\); (2) \(a^{(t)}\), the action type from \(\{\textsf{click},\textsf{text},\textsf{stop}\}\); (3) \(e^{(t)}\), the target element; and (4) \(p^{(t+1)}\), the page at step \(t+1\), which is the next page. We enforce \(a^{(T)}=\textsf{stop}\) and \(e^{(T)}=0\) for the terminal step. Let \(p_\mathrm{curr}\) denote the current page at the beginning of the action sequence generation, which provides the starting context (i.e., the HTML representation of the current interface and the current page ID) for task planning. The online action sequence generation process follows an iterative refinement approach. The \(\textsc{PromptGenerator}(\cdot)\) function constructs a prompt from the task description \(d\), knowledge graph \(K_i\), invalid step indices \(I\), and the current page \(p_\mathrm{curr}\).

Figure~\ref{fig:example-prompt} illustrates a concrete example of the prompt generated by the prompt generator. The prompt consists of both a fixed background description and dynamic content based on the current scenario. The knowledge graph transitions are represented as (X, A), indicating that interacting with element X will navigate to page A. The notation e\_i\_j within the brackets represents the element with id $j$ on page $p_{i}$.

The function \(\textsc{QueryActionSequence}(\cdot)\) generates a candidate sequence of actions from the prompt, while the function \(\textsc{Validator}(\cdot)\) examines the sequence against transition rules in \(K_i\) and returns invalid steps, if they exist. This process repeats until a valid sequence is produced or the maximum number of iterations is reached.
  \section{Implementation and Evaluation}

\subsection{Experimental Setups}

We implement GraphPilot using GPT-4o~\cite{hurst2024gpt} to evaluate its performance. We conduct comprehensive experiments on the DroidTask dataset~\cite{wen2024autodroid}, a benchmark specifically designed to assess the performance of end-to-end mobile task automation agents. DroidTask dataset comprises 158 high-level tasks extracted from 13 popular Android apps, providing a comprehensive evaluation suite that encompasses diverse mobile interaction scenarios and varying task complexities. We employ AutoDroid~\cite{wen2024autodroid} to extract the HTML representations of mobile app interfaces. Mind2Web~\cite{deng2023mind2web} and AutoDroid~\cite{wen2024autodroid} are selected as our baseline methods.

\subsection{Task Completion Rate}

We evaluate task completion performance using the task completion rate (TCR) metric, which is the percentage of tasks successfully completed out of the total number of evaluated tasks. A task is considered successfully completed when its generated sequence of actions exactly matches the ground truth sequence provided in the DroidTask dataset. This metric directly reflects the accuracy and effectiveness of a GUI agent.

\begin{figure*}[!htb]
    \centering
    \includegraphics[width=\linewidth]{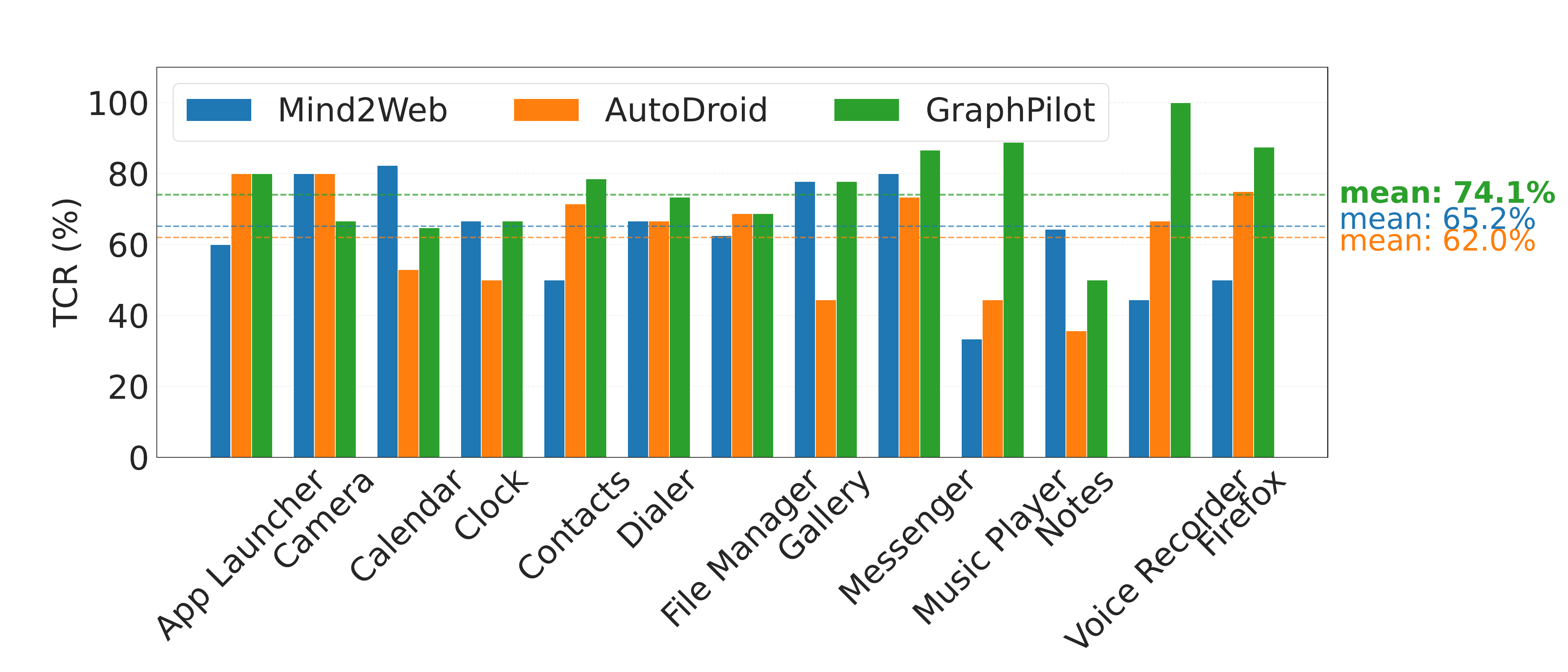}
    \captionof{figure}{TCR by apps and average TCR across different GUI agents.}
    \label{fig:tcr_apps}
\end{figure*}

We evaluate TCR for Mind2Web, AutoDroid, and GraphPilot. As summarized in Table~\ref{tab:tcr_comparison}, GraphPilot achieves the highest overall TCR of 74.1\% and consistently outperforms both baselines, leading Mind2Web by 8.9\% and AutoDroid by 12.1\%. The per-app breakdown in Figure~\ref{fig:tcr_apps} further shows that GraphPilot attains superior TCR on the majority of apps, indicating robust generalization across diverse tasks and interfaces. While performance varies across apps due to differences in task structure and GUI complexity, the advantage of GraphPilot remains stable.

\begin{center}
    \captionsetup{type=table}
    \captionof{table}{TCR comparison across different GUI agents.}
    \label{tab:tcr_comparison}
    \renewcommand{\arraystretch}{1}
    \begin{tabularx}{\linewidth}{@{}*{4}{>{\centering\arraybackslash}X}@{}}
        \toprule
        \textbf{Agent}      & Mind2Web & AutoDroid & \textbf{GraphPilot} \\
        \midrule
        \textbf{TCR (\%)}   & 65.2     & 62.0      & \textbf{74.1}       \\
        \bottomrule
    \end{tabularx}
\end{center}

To analyze the relationship between task complexity and performance, we measure task complexity by the number of actions required to complete the task. Figure~\ref{fig:num_of_operations} presents the TCR across different numbers of actions, revealing a downward trend as the number of actions increases for all GUI agents. Notably, GraphPilot almost always maintains a higher TCR than Mind2Web and AutoDroid across virtually all complexity levels, demonstrating its robustness in handling tasks of varying difficulty.

\begin{center}
    \includegraphics[width=\linewidth]{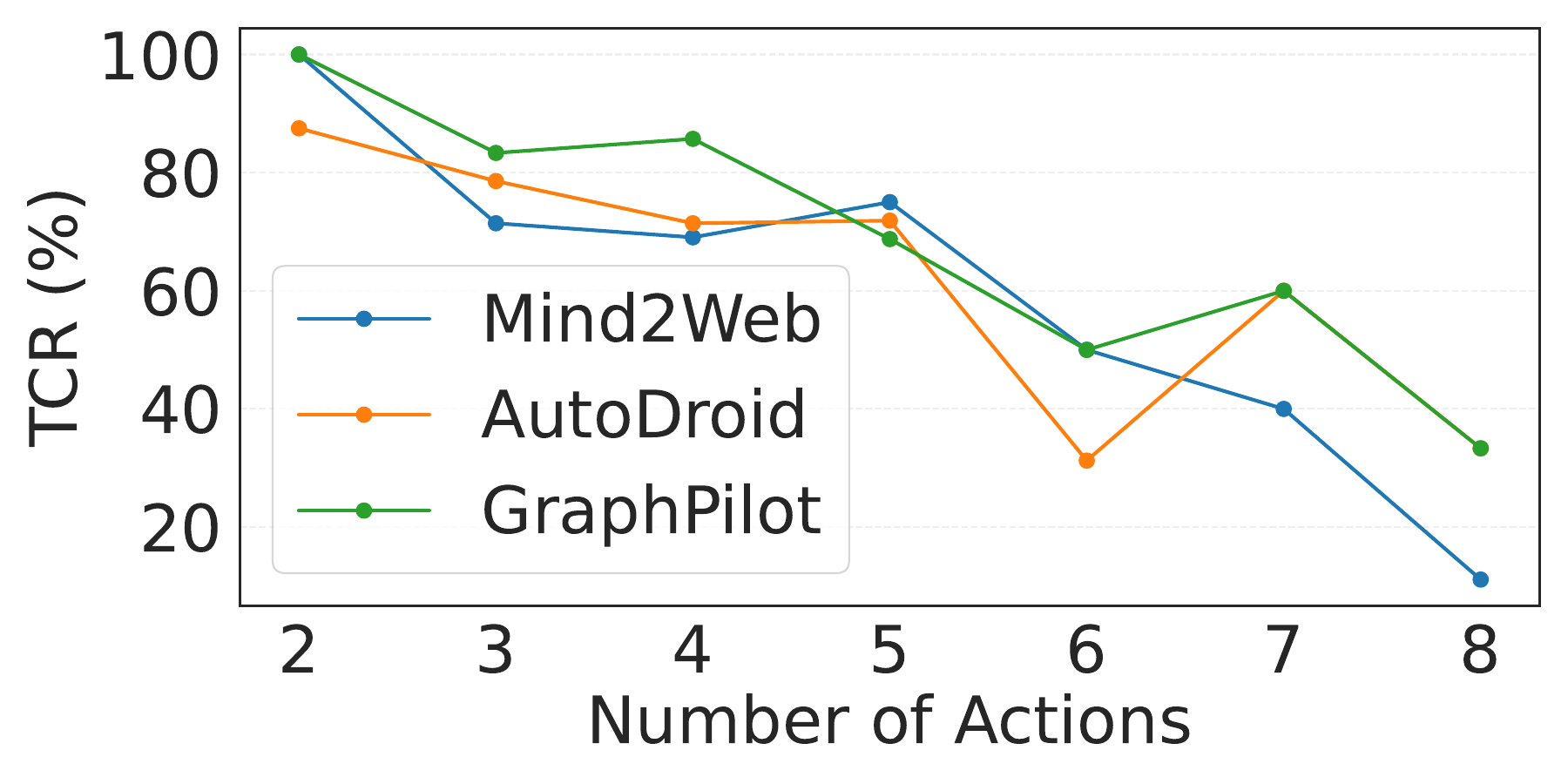}
    \captionof{figure}{TCR in different numbers of actions required to complete the task across different GUI agents.}
    \label{fig:num_of_operations}
\end{center}

\begin{figure*}[!htb]
    \centering
    \includegraphics[width=\linewidth]{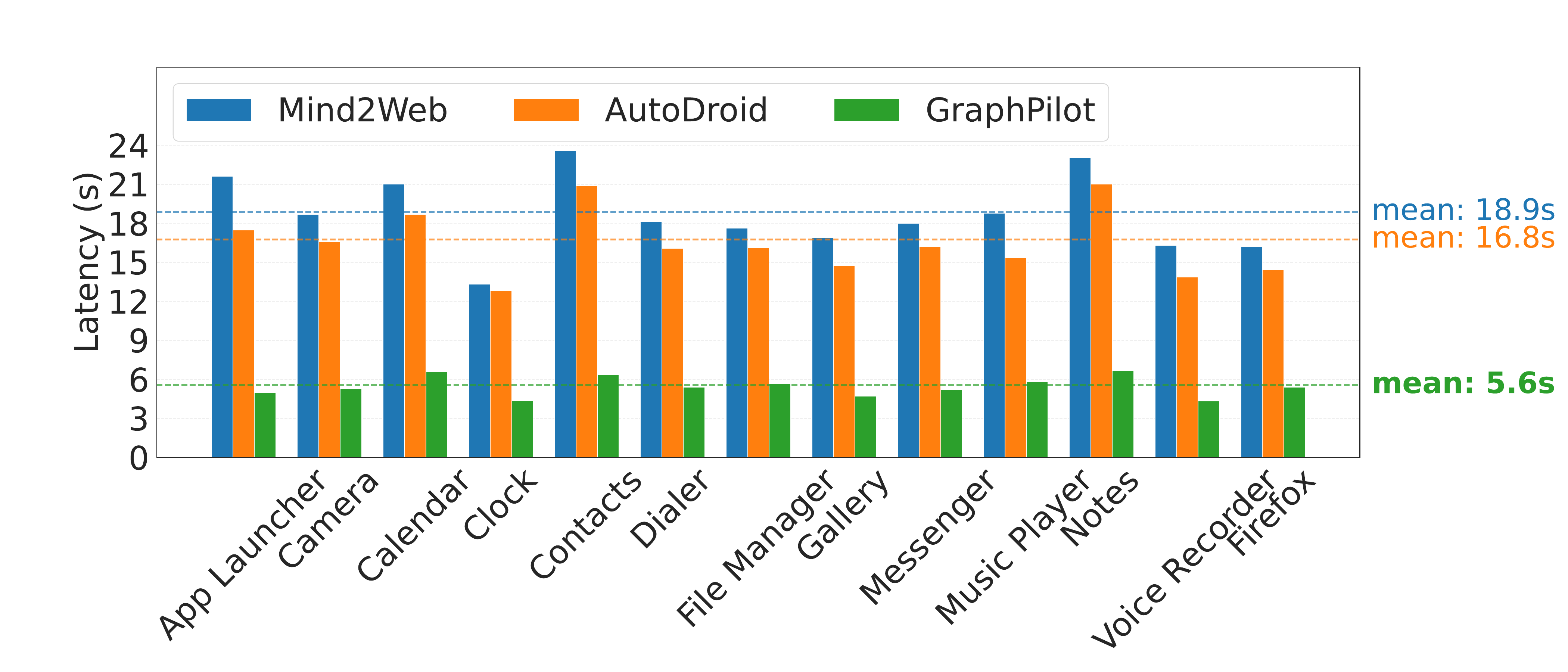}
    \caption{Latency comparison across different GUI agents.}
    \label{fig:latency}
\end{figure*}

\subsection{Latency}

We measure both the latency of LLM queries and the number of queries required for each task, as these represent the primary time costs for mobile GUI agents.
We compare the performance among Mind2Web, AutoDroid, and GraphPilot. The results are presented in Figure~\ref{fig:latency} and Table~\ref{tab:queries_comparison}.

\begin{center}
    \captionsetup{type=table}
    \captionof{table}{Comparison of the number of queries for each task across different GUI agents.}
    \label{tab:queries_comparison}
    \renewcommand{\arraystretch}{1}
    {\begin{tabularx}{\linewidth}{>{\hsize=0.6\hsize\centering\arraybackslash}X>{\hsize=1.4\hsize\centering\arraybackslash}X}
        \toprule
        \textbf{Agent}      & \textbf{Number of Queries Each Task} \\
        \midrule
        Mind2Web            & 4.54                          \\
        AutoDroid           & 4.54                          \\
        \textbf{GraphPilot} & \textbf{1.03}                 \\
        \bottomrule
    \end{tabularx}}
\end{center}

The results demonstrate that GraphPilot achieves significantly lower latency than Mind2Web and AutoDroid across all apps. In particular, GraphPilot reduces latency by 70.4\% compared to Mind2Web and by 66.7\% compared to AutoDroid. The reason is that GraphPilot requires substantially fewer LLM queries, representing a 77.3\% reduction in query frequency.
This improvement stems from GraphPilot's architectural design. This design typically requires only a single LLM query for each task to produce the complete sequence of actions. In contrast, agents like Mind2Web and AutoDroid must query the LLM for each action, which incurs additional latency.

\subsection{Ablation Study}

\begin{center}
\captionsetup{type=table}
    \captionof{table}{TCR of GraphPilot with different components removed.}
    \label{tab:ablation_gpt4o}
    \renewcommand{\arraystretch}{1}
    {\begin{tabularx}{\linewidth}{>{\centering\arraybackslash}X>{\centering\arraybackslash}X}
        \toprule
                                                         & \textbf{TCR (\%)}        \\
        \midrule
        \textbf{w/o transition rules}                    & 60.8 (13.3 $\downarrow$) \\
        \textbf{w/o validator}                           & 72.8 (-1.3 $\downarrow$) \\
        \textbf{w/o HTML request} & 71.5 (-2.5 $\downarrow$) \\
        \bottomrule
    \end{tabularx}}
\end{center}

To analyze the contribution of different components in GraphPilot, we conduct ablation experiments by removing components and evaluating their impact on TCR. The TCR drops by removing transition rules, validator, and dynamic HTML representation requests, shown in Table~\ref{tab:ablation_gpt4o}.

\subsubsection{Transition Rules}



The removal of transition rules underscores their critical importance. The numbers in parentheses represent the performance drop compared to the complete GraphPilot. Without transition rules, GraphPilot experiences substantial performance degradation (13.3\% drop). This significant decline highlights the key role of transition rules. They provide explicit navigation guidance during LLM reasoning, thereby preventing invalid interaction paths and ensuring more reliable task execution.

\subsubsection{Validator}


Removing the validator results in a slight performance degradation. While these drops are substantially smaller (1.3\% drop) than those observed when removing transition rules, they demonstrate that correcting invalid sequences generated by the LLM provides meaningful improvements to TCR. This suggests that, although the transition rules are clearly specified in the prompt, even an advanced LLM may occasionally produce action sequences that violate these rules. The validator serves as a quality assurance mechanism to ensure the reliability of execution.

\subsubsection{Dynamic HTML Representation Request}


Removing the dynamic HTML representation request causes a slight performance degradation (2.5\% drop). Because the information in the knowledge graph is rich enough for the LLM to provide the complete sequence of actions directly in most cases, dynamic requests are rarely made. This component serves as a fallback mechanism when uncertainty arises during reasoning and provides additional context when the knowledge graph is insufficient to support decision-making.

Overall, the experimental results demonstrate that transition rules, which represent the edges in the knowledge graph, contribute most significantly to TCR improvement among all evaluated components. This importance stems from their fundamental role in constraining the action space and providing explicit navigation guidance during LLM reasoning.

  \section{Discussion}

The knowledge graph can be naturally extended to multi-app collaborative scenarios. In such cases, nodes can represent functions of apps, pages, and elements, while edges can capture inter-app (element-to-page) and intra-app (element-to-app) transitions. This extension would enable GraphPilot to coordinate complex tasks that span multiple apps.

Currently, GraphPilot supports a limited action space consisting of $\textsf{click}$ and $\textsf{text}$ operations. Future work could expand this action space by supporting additional interaction types, such as $\textsf{drag}$, $\textsf{pinch}$, and $\textsf{long-press}$. This enhancement would enable GraphPilot to handle more complex mobile interactions and access a broader range of app functionalities.

Tasks can often be completed through multiple valid pathways. However, current evaluation methods require the output sequence to match the ground truth exactly, which may incorrectly classify valid alternative approaches as failures. Future research could explore more flexible evaluation approaches that assess task completion success rather than strict sequence matching, thereby allowing for recognition of alternative valid solution paths.

In the experiments, we assume that the information GraphPilot needs to perform tasks is available in the knowledge graph. GraphPilot currently relies on the integrity of the knowledge graph. If the knowledge graph lacks corresponding information, such as transition rules, the TCR of GraphPilot will be affected. Making GraphPilot more robust is our subsequent goal.
  \section{Conclusion}

We present GraphPilot, a mobile GUI automation agent that addresses the accuracy and latency limitations in conventional stepwise approaches by constructing a knowledge graph for each app. Evaluation on DroidTask demonstrates that GraphPilot outperforms Mind2Web and AutoDroid with improvements in TCR while dramatically reducing query latency and LLM queries. We believe that incorporating domain knowledge of apps into LLM prompts in an appropriate form will allow LLMs to handle GUI task automation more accurately and efficiently. This will enable the creation of practical personal AI assistants.
  \section*{Funding}

This work was supported in part by Guangdong S\&T Programme (Grant
No. 2024B0101040007); in part by the Guangdong Basic and Applied
Basic Research Foundation (No. 2023B1515120058); and the Program for Guangdong Introducing Innovative and Entrepreneurial Teams (No.
2017ZT07X355).
  \section*{Author Contributions}

Conceptualization, Mingxian YU and Xu CHEN; methodology, Mingxian YU; software, Mingxian YU; validation, Mingxian YU; formal analysis, Mingxian YU and Xu CHEN; investigation, Mingxian YU; resources, Xu CHEN; data curation, Mingxian YU; writing—original draft preparation, Mingxian YU; writing—review and editing, Siqi LUO; visualization, Mingxian YU and Siqi LUO; supervision, Siqi LUO and Xu CHEN; project administration, Xu CHEN; funding acquisition, Xu CHEN. All authors have read and agreed to the published version of the manuscript.

  \backmatter





  \section*{Conflict of Interest}
  \noindent All the authors declare that they have no conflict of interest.

  \section*{Data Available}
  \noindent The data and materials used in this study are available upon request from the corresponding author.




  \bibliography{ref}

\end{multicols}
\end{document}